\documentclass[aps,prl,twocolumn,superscriptaddress,showpacs,amsmath,amssymb]{revtex4-2}
\usepackage[utf8]{inputenc}
\usepackage{amsmath,amsthm}
\usepackage{graphicx}
\usepackage[colorlinks = true,
            linkcolor = blue,
            urlcolor  = blue,
            citecolor = blue,
            anchorcolor = blue]{hyperref}
\usepackage{braket}
\usepackage{xcolor}
\global\long\def\bra#1{\langle #1 |}
\global\long\def\ket#1{| #1 \rangle }
\global\long\def\kket#1{\| #1 \rangle\rangle}

\global\long\def\braket#1#2{\left\langle #1|#2\right\rangle }

\global\long\def\al{\alpha}
\global\long\def\be{\beta}
\global\long\def\ga{\gamma}
\global\long\def\de{\delta}
\global\long\def\De{\Delta}
\global\long\def\Ga{\Gamma}
\global\long\def\th{\theta}

\global\long\def\la{\lambda}
\global\long\def\ka{\kappa}
\global\long\def\si{\sigma}
\global\long\def\vfi{\varphi}

\global\long\def\tih{\tilde{h}}

\global\long\def\bbra#1{\langle \langle #1 \| }

\begin{document}

\title{Dissipative cooling towards phantom Bethe  states  in boundary driven XXZ spin chain
 }

\author{Vladislav Popkov}
\affiliation{Faculty of Mathematics and Physics, University of Ljubljana, Jadranska 19, SI-1000 Ljubljana, Slovenia}
\affiliation{Bergisches Universit\"at Wuppertal, Gauss Str. 20, D-42097 Wuppertal, Germany}
\author{Mario Salerno}
\affiliation{Dipartimento di Fisica ``E.R. Caianiello'', and INFN
- Gruppo Collegato di Salerno, Universitá di Salerno, Via Giovanni
Paolo II, 84084 Fisciano (SA), Italy}

\begin{abstract}
A dissipative method that allows to access family of phantom Bethe-states
(PBS) of boundary driven $XXZ$ spin chains, is introduced. The method consists
in  coupling  the ends of the open spin chain to suitable  dissipative
magnetic baths to force the edge spins to satisfy specific boundary
conditions necessary for the PBS existence.
Cumulative monotonous depopulation of  the non-chiral components of the density matrix
with growing dissipation amplitude is analogous to the depopulation of high-energy states in response to thermal cooling.
Compared to generic states, PBS have strong chirality, nontrivial topology and  carry  high spin currents.
\end{abstract}

\date{\today}
\maketitle

%\textbf{Vladislav Popkov $\rightarrow$ Mario Salerno,   Feb. 2, 2022}

\noindent\textbf{ Introduction.~} Dissipation needs not to be always destructive for quantum protocols but it can represent a resource for manipulating quantum systems. Dissipation alone
\cite{CiracDSE2009} or in combination with coherent dynamics
 \cite{2014ZanardiVenuti,BradlynGeometryResponseOfLindbladians2016,OscillationsZenoManifold2018,
DissGeneration2021,DarkStatesMultilevelAtoms2022,
GenerationMultiparticleEntanglementZeno2015,
Diehl2021,TopologyByDissipation2013,Kollath2016}, indeed,
can be used to create quantum  non-equilibrium stationary states (NESS) which are attractors of the dynamics and therefore are stable even in the presence of noise.

Most protocols, however, require a tailored set of operations used in pumping cycles  to target each specific state \cite{DissGeneration2021,ZollerQuantumSimulator2011}. If the protocols are
implemented by stationary control fields, they usually require sophisticated dissipations that make the NESS targeting  more complicated
\cite{KrausNonlocalJumpOper2008}.
%The latter obstacle is duet
This is due to the fact that the targeted NESS must be an eigenstate of the coherent part of the dynamics and a dark state for  all jump  operators in the dissipator \cite{Yamamoto2005,2017SHS-Linbdlad-Gunter}.

Here we demonstrate how to  generate a remarkable family of  NESS containing an arbitrary
number of qubits, employing simple boundary-localized dissipation, and manipulating just one parameter.  These are the phantom Bethe states (PBS), i.e. eigenstates of integrable  XXZ spin chains on special parameter manifolds  \cite{PhantomShort, PhantomLong, PhantomBetheAnsatz}, possessing unusual chiral and topological features.
%, e.g. carry high currents of magnetization.

The phenomenon is based on a subtle mechanism that makes (within the quantum Zeno limit \cite{Misra1977,ZenoStaticsExperimentalReview,QuantumZenoWineland1990,PascazioQZE2000}), a highly selective 'phantom' invariant subspace  the basin of attraction for the density matrix.
Consequently, the NESS responds in a singular resonant way to an increase of the dissipation strength in the vicinity of "phantom" manifolds, restricting the density matrix to states with chirality of the same sign and thus rendering  the  NESS chiral. The resonances become sharper as the  dissipation strength is increased and their number grows with the number of spins involved.

Dynamically, the ``freezing out" of  the non-chiral components of the density matrix with growing dissipation amplitude is analogous to the depopulation of high-energy states in response to thermal cooling.

The  ``dissipative cooling" method consists in  coupling  the ends of the open chain to dissipative baths of polarization constraining the first and the last spin  to relax to predefined pure qubit states that   satisfy  boundary conditions necessary for the PBS existence. We show that by changing the control parameter, i.e. the misfit azimuthal angle of the dissipatively-targeted boundary polarizations, it is possible to  thread  ``phantom" manifolds, passing from one chiral NESS to another one with  different topology (see Figs. \ref{Fig-JzN5}). To the best of our knowledge, a quantum protocol that allows to target the whole PBS family is presently lacking.

\begin{figure}[tbp]
\centerline{
\includegraphics[width=0.375\textwidth]{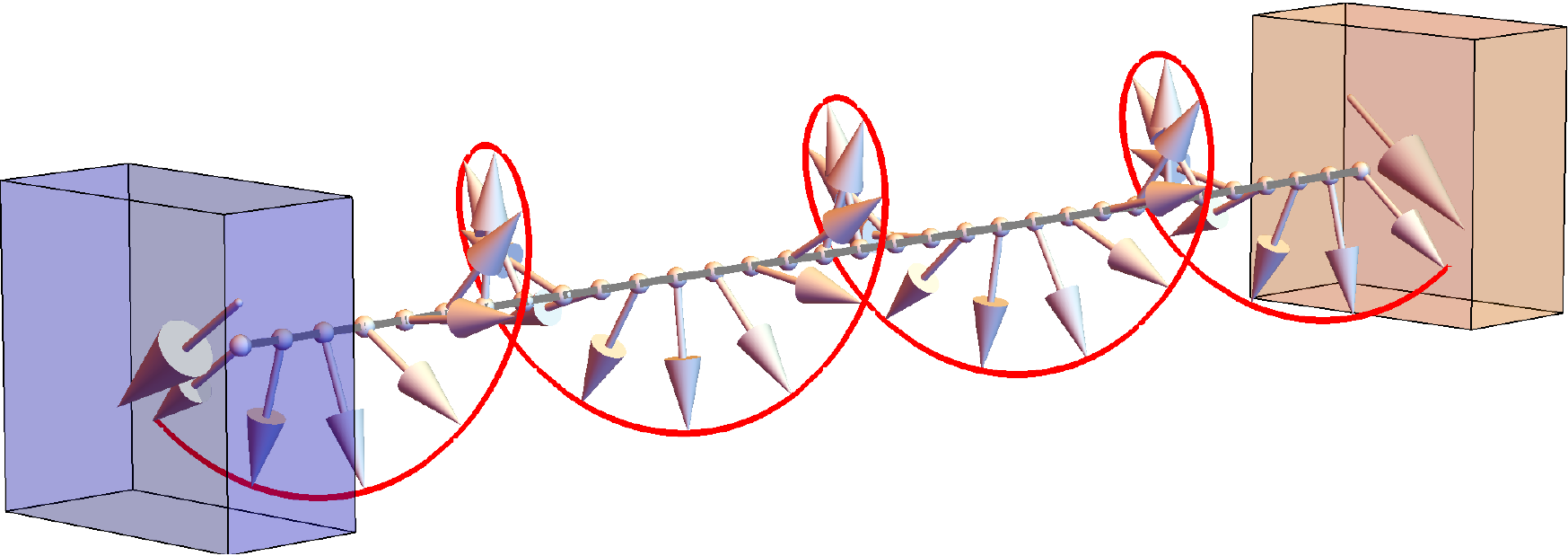}
}
\caption{
Model setup. An open XXZ spin chain is edge-coupled to dissipative fully polarizing baths (blue and pink  boxes). When baths's polarizations (big arrows inside boxes) match  Eq. (\ref{eq:Split}), phantom Bethe states carrying large spin currents come into existence. The figure shows a pure spin-helix state corresponding to case $M_{+}=0$ in Eq. (\ref{eq:Split}).
}
\label{Fig-Setup}
\end{figure}

The simplest states belonging to the phantom family are the spin-helix
states (SHS) (see top panel of Fig. \ref{Fig-SHS} for an example).
Recently, SHS were created and used as a sensitive tool to  measure the anisotropy in experiments on XXZ chains implemented via ultracold atoms \cite{JepsenPRX,Jepsen2021}. While in \cite{JepsenPRX,Jepsen2021}, the lifetime of the SHS is restricted by finite size effects, the stability of dissipatively created
SHS (and other PBS) is guaranteed as long as the boundary dissipation strength is kept sufficiently strong.

Note that while SHS are pure states, all other PBS are mixed states corresponding, in terms of the mean variation of the magnetization along the chain, to helices with a variable radius (see middle panel of Fig. \ref{Fig-SHS} for an example).
Quite interestingly, we find that in comparison to generic pure and mixed states of the open $XXZ$  chain, the  PBS have strong chirality and  carry relatively high spin currents (see Fig.\ref{Fig-JzN25}), a feature that can be of potential interest for future applications.

\noindent\textbf{Model setting and near Zeno limit relaxation.}
Our dissipative setup is schematically depicted in  Fig.~\ref{Fig-Setup}. An  $XXZ$ Heisenberg spin $\frac12$ chain of $N+2$ sites, numbered as $0,1,\ldots N+1$, is coupled to Lindbladian baths of polarizations at the ends (\cite{Petruccione, Misra1977}, see details  in \cite{SM}. We assume dissipative targeting of generic pure qubit states
$\rho_{L,R} =\left(I + \vec{n}_{L,R} \cdot \vec{\si}  \right)/2$ at sites $0$ and $N+1$, where $\vec{n}_{L}, \, \vec{n}_{R}$ are unit vectors of polarization.
The strength of the dissipation $\Ga$ is measured by the inverse time needed for edge spins to relax,  e.g.
$\rho_{L/R}(t) = \frac{I}{2}  + \frac12\vec{\si}_1 \vec{n}_{L/R} + O(e^{-\Ga t})$, if the coherent part ( XXZ spin chain Hamiltonian) is neglected.
If $\vec{n}_L \neq \vec{n}_R$, a nonequilibrium gradient is created and a spin current $j^z$ can flow,  typically obeying the
Fourier law  $j^z =O(1/N)$.

In (\ref{eq:Split}) we fix a resonance condition under which
steady currents can be increased up to maximally possible values, $j^z=O(1)$.   Using the criterion of  \cite{SpohnUniquenessNESS1977}
one finds that the NESS is always unique.

It was shown in \cite{2018ZenoDynamics} that close to quantum Zeno limit $\Ga \rightarrow \infty$ the relaxation to NESS undergoes a three-stage process, each stage occurring at different timescale.
On the shortest timescale: $t\leq O(1/\Ga)$, the boundary spins  relax towards their targeted states.
From this point on, the density matrix  is approximately given by
$\rho = \rho_L \otimes R(t) \otimes \rho_R$, where $R(t)= tr_{L,R} \rho(t)$ describes the time evolution of the the internal part of the system, which becomes approximately coherent again
\cite{2014ZanardiVenuti} and is
governed by the dissipation projected Hamiltonian $h_D$
given by \cite{2018ZenoDynamics}
\begin{align}
&h_D = \sum_{n=1}^{N-1} \vec{\si}_n \cdot \hat{J}\vec{\si}_{n+1} + \hat{J}\vec{n}_L\cdot \vec{ \si}_1 + \hat{J}\vec{n}_R\cdot \vec{ \si}_N,
\label{eq:hD}
\end{align}
with $\hat {J}= diag(1,1,\De)$. Note that the dissipation-projected Hamiltonian (\ref{eq:hD})  depends
on polarizations  of the baths through its boundary fields.

On the  intermediate timescale: $O(1/\Ga)\ll t\leq O(1)$, the reduced density matrix $R(t)$  acquires the approximate form
\begin{align}
&R(t) = \sum_\al P_\al\left( \tau \right) \ket{\al} \bra{\al}\ \ + O\left( \frac{1}{\Ga}\right) ,
\label{eq:R(t)}
\end{align}
where $\tau=\frac{t}{\Ga} $ and  $\ket{\al}$ are $h_D$ eigenstates, with  positive coefficients $P_\al$ significantly changing on long time scales
$\tau=O(1)$.
Finally, on the long time scale $O(1)\ll t\leq O(\Ga)$ the coefficients $P_\al(\tau) \rightarrow P_\al(\infty)\equiv P_\al$
relax to their stationary NESS values, by following
an  effective  Markov process
\begin{align}
&\frac{\partial P_\al(\tau)} {\partial \tau} = \sum_{\be \neq \al} w_{\be \al} P_\be(\tau) - \sum_{\be \neq \al} w_{\al \be} P_\al(\tau)
\label{eq:MP}
\end{align}
 with rates $w_{\al \be}\equiv w_{\al \rightarrow \be}$ given by
\cite{2018ZenoDynamics}
\begin{align}
&w_{\al \be} = |\bra{\be} g_L \ket {\al} |^2 +  |\bra{\be} g_R \ket {\al} |^2,
\label{eq:MPrates}
\end{align}
where $g_L,g_R$ are operators acting on the first and the last spin respectively, given by
\begin{align}
&g_L =g(\vec{n}_L) \otimes I^{\otimes_{N-1}}, \quad
g_R = I^{\otimes_{N-1}} \otimes g(\vec{n}_R),\\
&g\left( \vec{n}(\th,\vfi)\right) = \hat{J} \left(
\vec{n}(\frac{\pi}{2}-\th,\vfi+ \pi) - i  \vec{n}(\frac{\pi}{2},\vfi+ \frac{\pi}{2})
\right)\cdot \vec{\si},
\end{align}
where $\th,\vfi$ are spherical coordinates of a unit vector.
The final state, the NESS, in the Zeno limit has the form
\begin{align}
&\rho_{NESS}^{Zeno} =\rho_L \otimes \left( \sum_\al P_{\al} \ket{\al} \bra{\al} \right) \otimes \rho_R,
\label{eq:rhoNESS}
\end{align}
where $P_{\al}$ is time-independent solution of (\ref{eq:MP}), and
$\ket{\al}$ are eigenstates of $h_D$ (\ref{eq:hD}).

\noindent\textbf{Phantom Bethe states.}
Phantom Bethe states are eigenstates of the Hamiltonian (\ref{eq:hD}) that have
exceptional chirality and  correspond to special manifolds of the boundary fields
in (\ref{eq:hD}) $\hat{J}\vec{n}_L (\th_L,\vfi_L)$,  $\hat{J}\vec{n}_R (\th_R,\vfi_R)$, with $\th,\vfi$ given by
\cite{PhantomShort,PhantomLong}
\begin{align}
\th_L = \th_R, \quad \vfi_R - \vfi_L = (N+1 -2 M_{+})\ga,  \label{eq:Split}
\end{align}
where $M_{+}=0,1, \ldots N+1$ \cite{PhantomCriterium} and $\ga$ parametrizing the $XXZ$ model anisotropy $\De$
\begin{align}
& \De= \cos \ga. \label{Delta}
\end{align}
The substitution $M_{+} \rightarrow M_{-} \equiv N+1-M_{+}$ in (\ref{eq:Split}) leads to the physical setup with the
 opposite boundary gradient, $\vfi_R - \vfi_L \rightarrow - (\vfi_R- \vfi_L)$,
and consequently flips the steady current $j^z \rightarrow -j^z$.
The respective NESSs are related via   the left-right reflection $\vfi_L \leftrightarrow \vfi_R $
and subsequent rotation in $XY$-plane, see  \cite{SM}.
Using this property we restrict
the $M_{+}$ range to the  $M_{+}=0,1, \ldots [\frac{N+1}{2}]$.

For fixed $M\equiv M_{+}$ in this range all eigenstates $\ket{\al}$ of $h_D$ (\ref{eq:hD}) which
determine Zeno NESS (\ref{eq:rhoNESS}) split into
two chiral families, $\{ \ket{\al_+} \}$,  $\{ \ket{\al_{-}} \}$  characterized by
the so-called phantom  Bethe roots \cite{PhantomShort}. All  eigenstates from each family share common
chiral properties  \cite{PhantomLong}. Introducing the function $b(n,m)=\sum_{k=n}^m \binom{N}{k}$,
 the number of eigenstates $d_{+},d_{-}$ in  $\{ \ket{\al_+} \}, \ \{ \ket{\al_-} \}$ is given
by $d_{\pm}=b(0,M_{\pm})$.
In addition, in our case (\ref{eq:Split}) a smaller invariant subfamily
$\{ \ket{\al_+^{(1)}} \} \in \{ \ket{\al_+} \}$
exists \cite{SM}, yielding further splitting $\{ \ket{\al_+} \} = \{ \ket{\al_+^{(1)}} \} \oplus   \{ \ket{\al_+^{(2)}} \}$
 where $d_{+}^{(1)}=b(M-1,M) =\binom{N+1}{M}$,  $d_{+}^{(2)}= b(0,M-2)$.

According to the above,
the reduced density matrix $R$ in (\ref{eq:R(t)}) on ``phantom" manifolds (\ref{eq:Split}) splits as
\begin{align}
&R \approx \sum_{\ka=1}^2
\sum_{\al_{+}^{(\ka)} = 1}^{d_{+}^{(\ka)} } P_{\al+}^{(\ka)}\left( \tau \right) \ket{\al_{+}^{(k)}} \bra{\al_{+}^{(k)}} +
\sum_{\al_{-} = 1}^{d_{-} } P_{\al-}\left( \tau \right) \ket{\al_{-}} \bra{\al_{-}}.
\label{eq:Rsplitted(t)}
\end{align}

The sum (\ref{eq:Rsplitted(t)}) contains projectors on states with opposite chiralities and is generically approximately neutral. The time evolution
obeys the effective Markov process (\ref{eq:MP}) i.e. depends on rates $w_{\be \de}$ exclusively. Analyzing the rates \cite{SM} we find
a remarkable property: all  the rates
\begin{align}
&w_{\al_{+}^{(1)} \rightarrow  \al_{-} }
=w_{\al_{+}^{(1)} \rightarrow  \al_{+}^{(2)} }=0
\label{eq:zero-rates}
\end{align}
vanish, while  generic $w_{\be \rightarrow  \al_{+}^{(1)} }$ remain finite.
Thus, the subfamily $ \{ \ket{\al_+^{(1)}} \} $ becomes an adsorbing basin of the Markov process (\ref{eq:MP}) \cite{GamblerRuin2020}
resulting in depopulation of all other states with time, and
leading to the NESS of the form
\begin{align}
&\rho_{NESS}^{Phantom}(M) =\rho_L \otimes \left( \sum_{\al_{+}^{(1)} =1}^{\binom{N+1}{M}} P_{\al_ {+}^{(1)} } \ket{\al_{+}^{(1)} } \bra{\al_{+}^{(1)} } \right) \otimes \rho_R.
\label{eq:rhoNESS-Phantom}
\end{align}
All eigenstates $\ket{\al_{+}^{(1)} }$ in  (\ref{eq:rhoNESS-Phantom}) are phantom Bethe eigenstates
\cite{PhantomShort} of the same chirality and the chirality gets more pronounced for small
 $M$. For $M=0$,  the sum in  (\ref{eq:rhoNESS-Phantom}) contains  just one term, a
projector on the  spin-helix state
(SHS) (\ref{eq:SHS})
\begin{align}
&\Psi_{SHS}(\ga)= \bigotimes_{k=1}^{N} \binom{\cos \frac{\th_L}{2} }{\sin \frac{\th_L}{2} \ e^{ i k\ga + i \vfi_L}},
\label{eq:SHS}
\end{align}
visualized in Fig.~\ref{Fig-Setup}, characterized by a large current of magnetization
 \begin{align}
& j_{SHS}=\langle 2 (\si_n^x \si_{n+1}^y - \si_n^y \si_{n+1}^x) \rangle_{SHS}= 2\sin^2 \th_L \ \sin\ga.
\label{eq:jSHS}
\end{align}
A possibility to target spin-helix state (\ref{eq:SHS})  dissipatively was also noted in  previous studies
\cite{2017TargetingNoncommutive,2017SHS-Linbdlad-Gunter,2017SHS-Carlo}.

For $M>0$ the  ideal helix (\ref{eq:SHS}) gets blurred but the NESS (\ref{eq:rhoNESS-Phantom}) remains chiral. For $M=1$ the current of magnetization $\bra{\al_{+}} \hat{j}^z \ket{\al_+}$ averaged over  states $\ket{\al_+}$ is of the order $\langle j^z  (M)\rangle \approx j_{SHS} (1-2/N)$,  while
for arbitrary $M<N/2$, estimates yield $\langle j^z(M) \rangle \approx j_{SHS} (1-2M/N)$  \cite{PhantomLong}.

From the above result we predict the existence of chiral Zeno NESS with unusually high  magnetization current at phantom Bethe manifolds (\ref{eq:Split}).
\begin{figure}[tbp]
\centerline{
\includegraphics[width=0.46\textwidth]{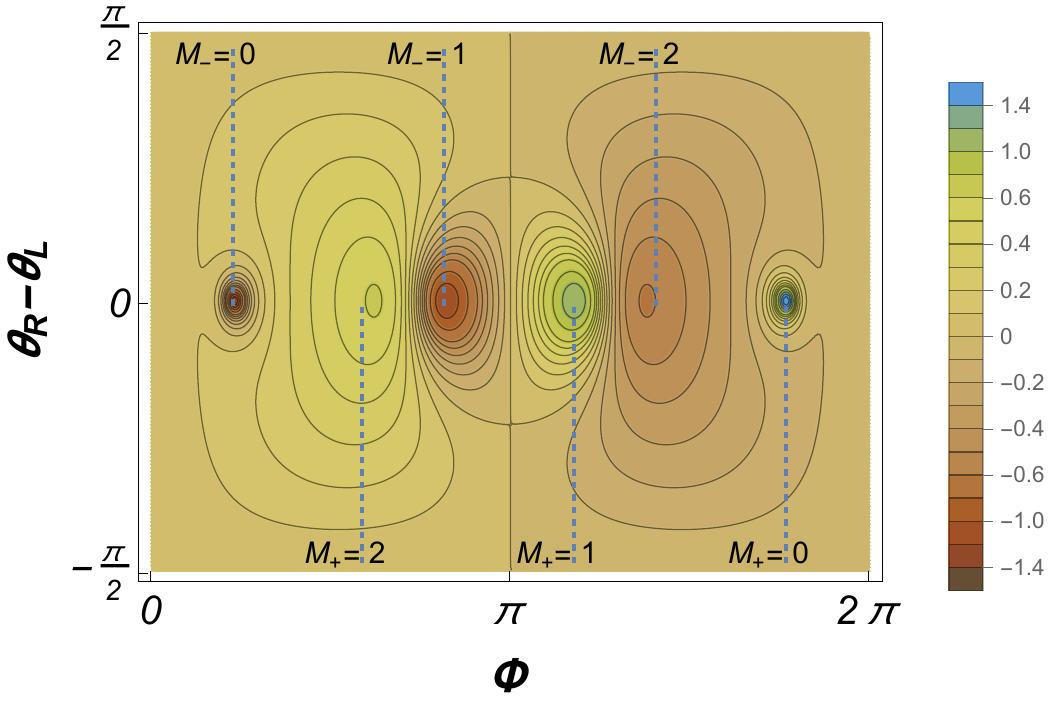}}
\vspace{-2mm}
\centerline
{\includegraphics[width=0.375\textwidth]{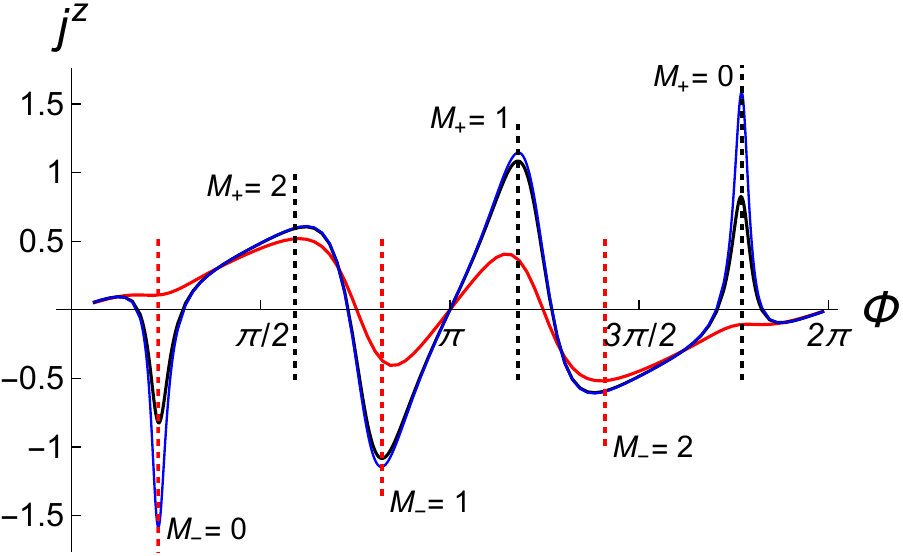}
}
\caption{\textbf{Top Panel}: Steady magnetization current $j^z$ versus  boundary
misfits of azimuthal angle 
$\Phi= \vfi_R - \vfi_L$ and polar angle
 $\th_R-\th_L$, in Zeno limit, for $N=5, \De=0.6, \th_L=\pi/2$,  and $\vfi_L=0$ The targeted spin
polarizations are  $\vec{n}_L=(1,0,0),
\vec{n}_R=(\sin \th_R \cos \Phi, \sin \th_R \sin \Phi ,\cos \th_R)$.
\textbf{Bottom Panel}: cut of the $j^z$ surface at $\th_R-\th_L=0$,
for different dissipation strengths $\Ga =  20,100,1000$ (more spiky functions correspond to
increasing $\Ga$.
Vertical black and red dotted lines indicate the misfit angles $\Phi= \pm(N+1-2M)\ga$.
 }
\label{Fig-JzN5}
\end{figure}

\noindent \textbf{Numerical results.~}
To check our predictions we study  NESS magnetization current  for fixed  anisotropy $\De$ and
varying boundary gradient, for systems of size $4 \leq N \leq 30$. We
use exact numerical diagonalization for small chains \cite{SPPLA2013} and Matrix product ansatz
for NESS in the Zeno limit \cite{2020ZenoPRL,MPA2021} for large chains.
Already for $N=5$ we find all  $j^z$ peak positions at predicted points  (\ref{eq:Split}) and their vicinity, see Figs.~\ref{Fig-JzN5}. Namely, all points   (\ref{eq:Split}) with $M_\pm=0,1,2$
correspond to peaks of various amplitudes ($j^z=0$ for   $M_{+}=(N+1)/2=3$,
since it corresponds to zero misfit angle $\Phi= \pm(N+1-2M_{+})\ga=0$
and hence to absence of  the boundary gradient $\vec{n}_L = \vec{n}_R$).

For larger chains (Fig.~\ref{Fig-JzN25}) the agreement and the phenomenon becomes striking: most $j^z$ peaks appear as sharp resonances centered at phantom Bethe states manifolds (\ref{eq:Split}), on top of a background
with $j^z = O(1/N)$. Notice that the empty black and red circles in (Fig.~\ref{Fig-JzN25}) show the dependence of the $j^z$ peaks on $M$ (top horizontal-axis).
\begin{figure}[tbp]
\centerline{
\includegraphics[width=0.5\textwidth]{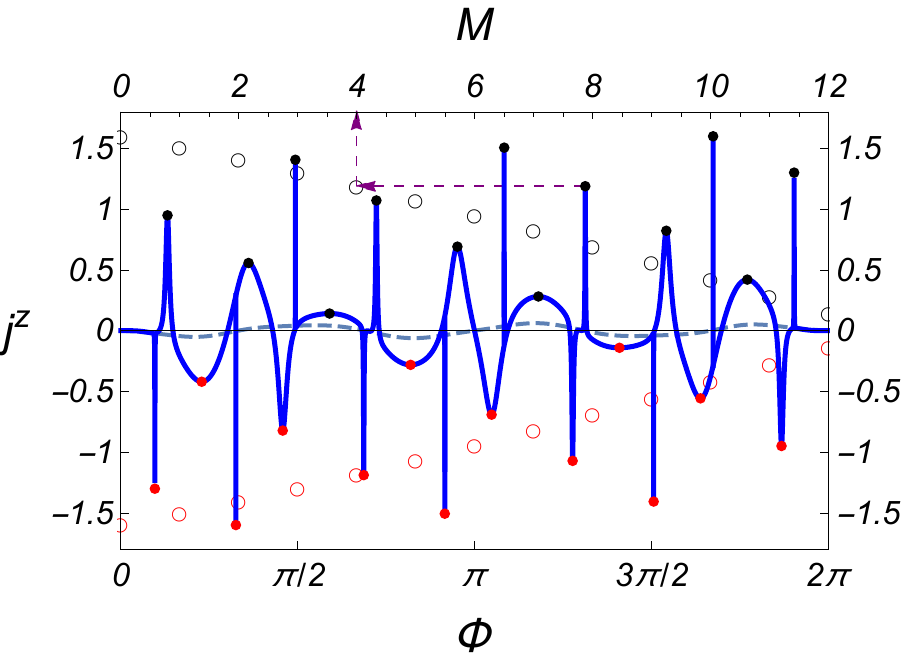}}
\caption{Steady magnetization current $j^z$ (blue curve) versus $XY$-plane misfit angle $\Phi= \vfi_R - \vfi_L$ (bottom horizontal-axis)
in the Zeno limit $\Ga \rightarrow \infty$,
computed using method in \cite{2020ZenoPRL,MPA2021}.  Parameters are: $N=25$, $\De=0.6, \th=\pi/2$.  $\Phi$ coordinates of black/red dots correspond to angles $\Phi_c(M)=\vfi_R-\vfi_L$ with
$M=0,1,\ldots , N+1$ in (\ref{eq:Split}).
The dashed blue line shows $j_z$ versus $\Phi$ for a  chain  with an additional boundary misfit $\pi/6$ in the polar angle: $\vec{n}_L=(1,0,0)$,
$\vec{n}_R=(\frac{\sqrt{3}}{2}\cos \Phi,\frac{\sqrt{3}}{2}\sin \Phi ,\frac{1}{2})$.
The empty black and red circles referring to the top horizontal-axis permit to identify the values of $M$ associated to each $j^z$ peak, as indicated by the purple dashed lines with arrows (find the corresponding open circle of equal amplitude and read the $M$ value on the top-horizontal axis).
}
\label{Fig-JzN25}
\end{figure}

The role of the parameter $M$ determining the NESS rank in (\ref{eq:rhoNESS-Phantom}) via $r (M)= \binom{N+1}{M}$ deserves  special discussion.
The highest and sharpest of all $j^z$ peaks   always corresponds to
$M_\pm=0$, see Figs.~\ref{Fig-JzN5}, \ref{Fig-JzN25}, i.e. pure Zeno NESS, with magnetization winding in a perfect helix around the $z$ axis, see (\ref{eq:SHS}) and Fig.~\ref{Fig-SHS}.
For $M>0$,  perfect chirality is lost: the basis of ``phantom" manifold for $M=1$ consists of $2$ disjointed  helix pieces  separated by a kink at some position $n$; phantom Bethe states are linear combinations  of basis states for all possible kink positions \cite{PhantomLong}. Resulting NESS magnetization profile is a distorted helix with  variable radius, see mid-panel of Fig.~\ref{Fig-SHS}.
Basis states for arbitrary $M$ contain $M$ kinks  \cite{PhantomLong}, introducing more  helix imperfections.  For large $M$  NESS becomes a mixture of exponentially large number of states.
E.g. the peak with $M=8$ in Fig.~\ref{Fig-JzN25} has $\binom{26}{8}> 1.5*10^6$ contributing states, about 5$\%$ of full Hilbert space containing  $2^{25}$ states. Nevertheless, due to the similar chiral properties of all contributing components, the respective NESS is chiral.

\begin{figure}[tbp]
\vspace{-0mm}
\centerline{
\includegraphics[width=0.36\textwidth]{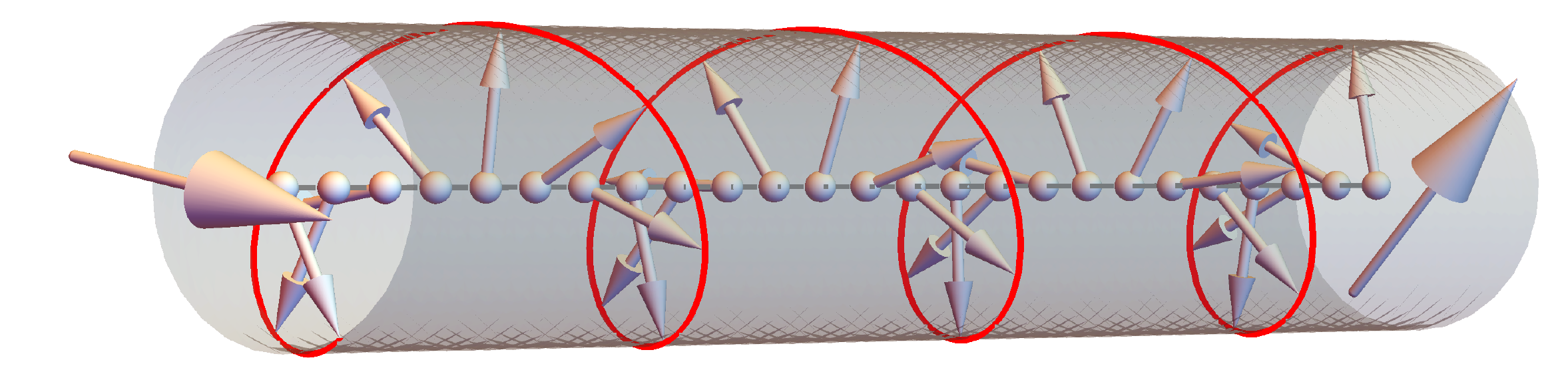}
}
\vspace{6mm}
\centerline{
\includegraphics[width=0.36\textwidth]{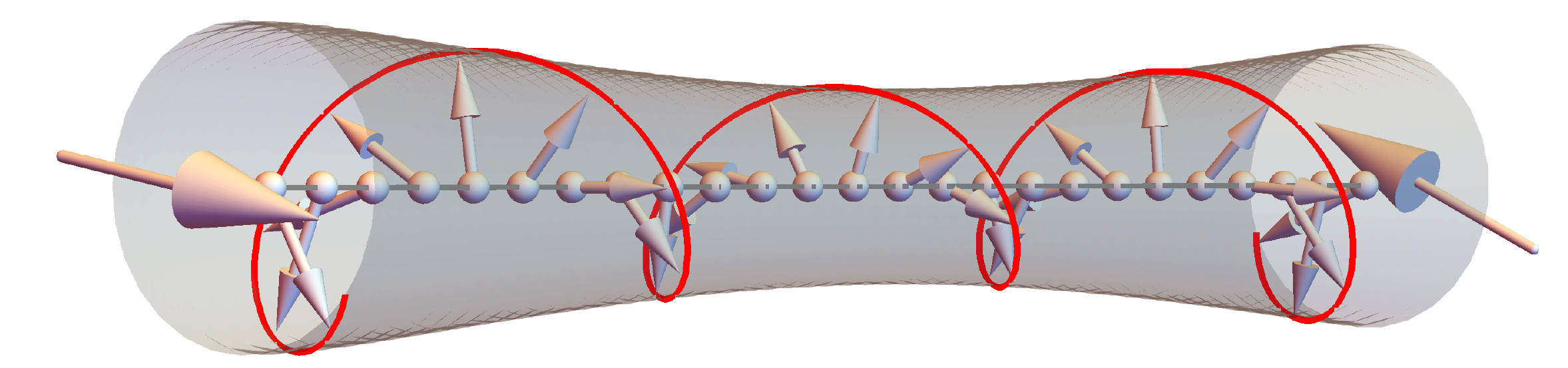}}
\vspace{6mm}
\centerline{
\includegraphics[width=0.3\textwidth]{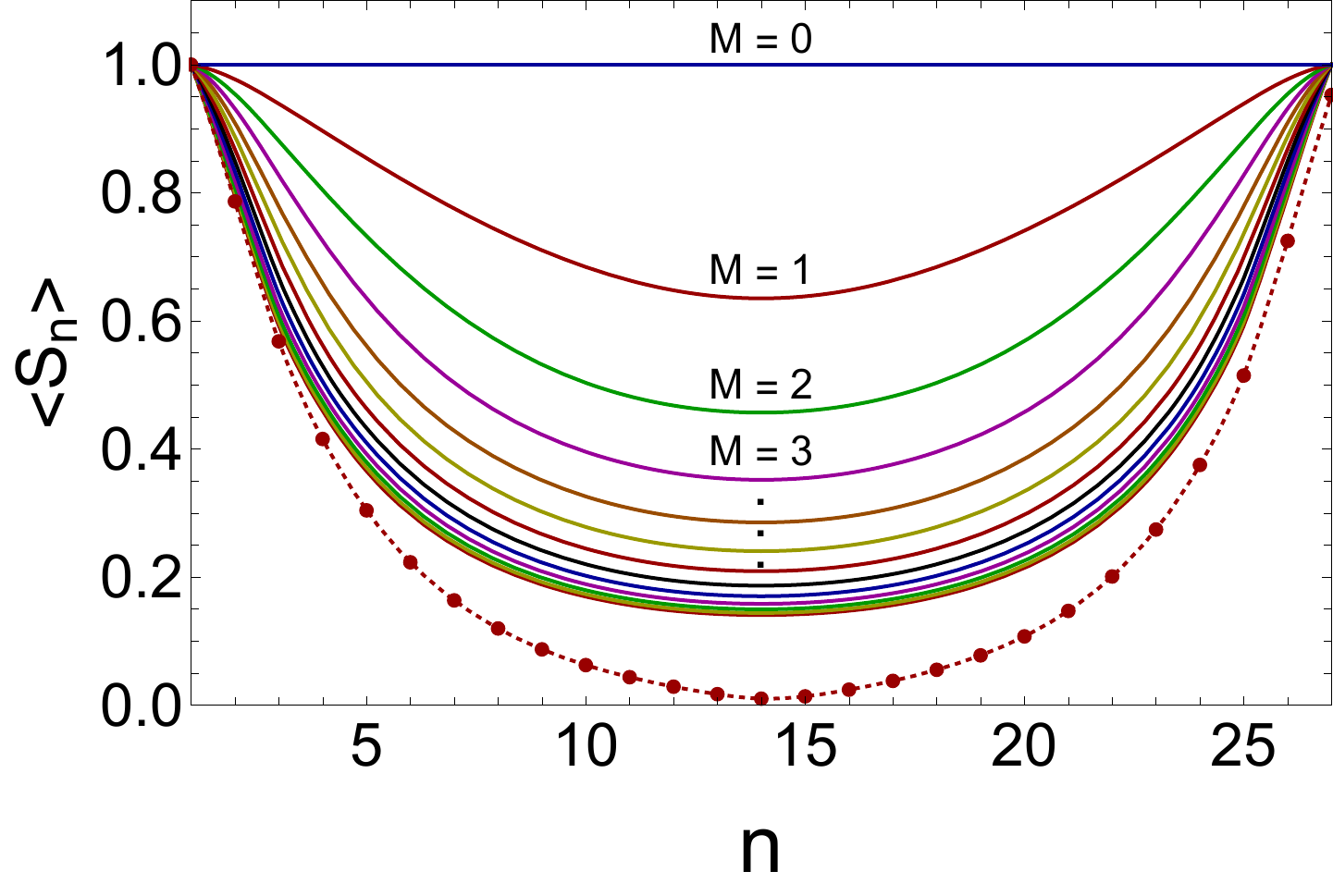}
}
\caption{
{\bf Top and middle panels}. PBS of the XXZ open chain with $N=25$ and for cases $M=0$ (top panel) and  $M=1$ (middle panel) in  Eq.(\ref{eq:Split}).  The $z$-axis is directed along the chain while the left big arrow  points in the positive $x$-direction.
{\bf Bottom panel:} Variation of the averaged spin (helix radius) along the chain for $M=0,1,...,12$ (curves from top to bottom, respectively). The dotted curve at the bottom, depicted for comparison, refers to a generic NESS out of the phantom manifold. All parameters are fixed as in Fig.~\ref{Fig-JzN25}.
}
\label{Fig-SHS}
\end{figure}
%

%Thus, with simple variation of boundary misfit, one can obtain  mixed states with chiral properties.

It is also worth to note that, due to larger number of contributing states, peaks with larger $M$ are less sharp and have smaller  amplitude, but, on the other side, the NESS gets more
stable with respect to perturbations (boundary misfits, lowering dissipation, etc.). This effect can be seen  by comparing the behavior of peaks at increasing $\th_R-\th_L$ misfit (upper panel of Fig.~\ref{Fig-JzN5}) and for different $\Ga$ values  (the bottom panel).

In particular, as $\Ga$ decreases,  the
near Zeno limit description of an effectively coherent evolution (\ref{eq:R(t)}) becomes invalid, and the peaks gradually smear out, the sharper peaks first.

In contrast, for parameters chosen away from ``phantom" manifolds, variations of
$\Ga$ do not lead to any  drastic effects  (data not shown), especially for large enough $N$. The reason is, beyond certain characteristic  value $\Ga>\Ga_{ch}(N)=O(1/N)$, effective quantum Zeno regime sets in.

\noindent \textbf{Conclusion.} We demonstrated that the spikes of the steady magnetization current in open $XXZ$ spin chains with dissipatively created  boundary gradient are due to the existence of special manifolds (\ref{eq:Split}) where  phantom Bethe roots
solutions  of the Bethe Ansatz equations exist.
The mechanism by which  the  PBS can be physically accessed was identified to be
strong dissipation, which drives the system towards a  chiral invariant subspace. The
 gradual depopulation of the non-chiral states resembles the  depopulation of  highly energetic states in a quantum system coupled to a cold thermal bath.
The amplitude of dissipation plays the role of inverse temperature and a chiral subset of states plays the role of a subset of low-energy states close to the ground state of the system. We showed that by varying the system size, the bulk anisotropy and the boundary dissipative driving
one can manipulate a number of peaks of the magnetization current,  distance between the peaks in the parameter space, and their magnitudes. All the resulting stationary states are easily distinguishable by the value of the carried spin current, and by their nontrivial topology \cite{WindingSpinHelices2019}.

We expect the dissipative cooling approach considered  in this paper to be effective also for other open quantum many body systems that are integrable via the off-diagonal Bethe ansatz method.

\begin{acknowledgments}
VP acknowledges financial support from the European Research Council through the advanced grant No. 694544—OMNES, and from the Deutsche Forschungsgemeinschaft through the DFG projects KL 645/20-1, KL 645/20-2. VP thanks the Department of Physics "E.R.Caianiello" for hospitality and for short visits partial supports (FARB 2018 - 2019) during which the work was completed.
\end{acknowledgments}

%\bibliographystyle{apsrev4-2}
%\bibliography{LindbladPhantomRef}

\begin{thebibliography}{35}
\bibitem [{Pha()}]{PhantomCriterium}%
  \BibitemOpen
  \href@noop {} {}\bibinfo {note} {Phantom Bethe roots criterium for open $XXZ$
  spin chain with boundary fields is given by Eq.(17) of \cite{PhantomLong}.
  For our special case $h_D$ (\ref{eq:hD}) we set $\al_{+}= \al_{-}=-\eta=-i
  \ga$, $\be_{+}+\be_{-}=0$, $\th_{-}= i (\pi-\vfi_L)$, $\th_{+}= i
  (\pi-\vfi_R)$, $1/\cosh (\be_{-})=\sin \th_L$, $\tanh (\be_{-})=-\cos \th_L$,
  leading to Eq.(\ref{eq:Split}), with $M=0,\ldots, N-1$. Another choice of
  parameters $\al_{+}= \al_{-}=\eta=i \ga$, $\be_{+}+\be_{-}=0$, $\th_{-}= -i
  \vfi_L$, $\th_{+}= -i \vfi_R$, $1/\cosh (\be_{-})=\sin \th_L$, $\tanh
  (\be_{-})=\cos \th_L$ leads to the same $h_D$. Inserting it into Eq.(17) of
  \cite{PhantomLong} leads to the same Eq.(\ref{eq:Split}) with $M \rightarrow
  M+2$, thus enlarging the $M$ range in (\ref{eq:Split}) to values $M=0,\ldots,
  N+1$.}\BibitemShut {Stop}%
\end{thebibliography}

%apsrev4-2.bst 2019-01-14 (MD) hand-edited version of apsrev4-1.bst
%Control: key (0)
%Control: author (72) initials jnrlst
%Control: editor formatted (1) identically to author
%Control: production of article title (-1) disabled
%Control: page (0) single
%Control: year (1) truncated
%Control: production of eprint (0) enabled
%

\clearpage

\setcounter{table}{0}
\renewcommand{\thetable}{S\arabic{table}}%
\setcounter{figure}{0}
\renewcommand{\thefigure}{SM\arabic{figure}}%
\setcounter{equation}{0}
\renewcommand{\theequation}{S\arabic{equation}}%
\setcounter{page}{1}
\renewcommand{\thepage}{SM-\arabic{page}}%
\setcounter{secnumdepth}{3}
\setcounter{section}{0}
\renewcommand{\thesection}{\arabic{section}}%
\setcounter{subsection}{0}
\renewcommand{\thesubsection}{\arabic{section}.\arabic{subsection}}%
\renewcommand{\thesection}{S-\Roman{section}}

\onecolumngrid

\global\long\def\no{\nonumber}

\begin{center}\Large{
		\textit{Supplemental Material for}\\
		\medskip
		
		\textbf{Dissipative cooling  towards  phantom Bethe  states \\ in boundary driven XXZ spin chain}\\
		
		\medskip
		by Vladislav Popkov and Mario Salerno
	}
\end{center}

{
\hypersetup{linkcolor=black}
%\tableofcontents
}

%\addcontentsline{toc}{section}{Overview}

\bigskip

This Supplemental Material contains four sections organized as follows. In \ref{S-I} we collect some useful standard definitions. In \ref{S-Ib} we derive symmetry properties of the Lindblad equation.
In  \ref{S-II} we prove the existence of invariant subspaces at Bethe manifolds (\ref{eq:Split}).
In  \ref{S-III}
we analyze the effective Markov process (\ref{eq:MP}).

\section{Lindblad Master equation and system evolution near Zeno limit}
\label{S-I}

A density matrix of a generic quantum system with dissipation satisfies, under standard assumptions
\cite{Petruccione, Misra1977}, a Lindblad Master equation (LME),
\begin{align}
&\frac{\partial \rho(t)} {\partial t} = - i [H_s,\rho] + \Ga {\cal D}[\rho]
\label{eq:LME}\\
&{\cal D}[\rho] = \sum_\ga {\cal D}_{L_{\ga}}[\rho] = \sum_\ga  2L_\ga \rho L_\ga^\dagger -  (L_\ga^\dagger L_\ga \rho + \rho  L_\ga^\dagger L_\ga ),
\nonumber
\end{align}
where $H_s^\dagger = H_s $ is a coherent part,
${\cal D}[\rho]$ is the dissipator,
and
$L_\ga$ are Lindblad operators, describing the interaction with the environment.
In our model $H_s$ is taken as the Heisenberg XXZ Hamiltonian with $z$ anisotropy $J_z/J_x = J_z/J_y= \De$:
\begin{align}
&H_s =\sum_{n=0}^N \si_n^x   \si_{n+1}^x + \si_n^y   \si_{n+1}^y + \De \si_n^z   \si_{n+1}^z
= \sum_{n=0}^N \vec{\si}_n \cdot \hat{J}\vec{\si}_{n+1},
\label{eq:Hs}
\end{align}
with  $\hat {J}= diag(1,1,\De)$. The chain consists of $N+2$ sites,
numbered as $0,1,\ldots N+1$, and is coupled at the edges, i.e. at sites 0 and N+1, to magnetic reservoirs modeled by Lindblad operators,
$L_i$, $i=L,R$ for left edge ($L \equiv site\  0$) and right edge ($R \equiv site\  N+1$), respectively,
constraining boundary spins to relax to generic pure qubit states of the form
\begin{align}
&\rho_{L,R} =\frac{1}{2}\left( I +  \vec{n}_{L,R} \cdot \vec{\si}  \right),
\label{eq:polarization}
\end{align}
with  $\vec{n}_{L}, \, \vec{n}_{R}$ unit vectors of polarization, parametrized by spherical coordinates angles $0\leq \th_{L},\th_{R} \leq \pi$ and  $0\leq \vfi_{L},\vfi_{R} < 2\pi$.

The explicit form of $L_{L,R}$ can be obtained by observing that the targeting of a pure qubit state $\ket{\psi} \bra {\psi}$ at a single site, is achieved by means of operators $L$ of the form $L =\ket{\psi} \bra {\psi^\perp}$,  with $\braket{\psi^\perp}{\psi}=0$, so that the dissipator ${\cal D}_L[\rho]$ automatically satisfies ${\cal D}_L[\ket{\psi} \bra {\psi}]=0$.
Observing  that the normalization of the states $\ket{\psi},\ket{\psi^\perp}$ entails
$tr( L^\dagger L) =1$ and that the  linear map ${\cal D}_L[\rho]$ has four eigenmodes with eigenvalues $0,-1,-1,-2$, one can write a formal  solution, valid for large times $t$, of the Eq. (\ref{eq:LME}) with $H_s \rightarrow 0$ and with a single operator $L=\ket{\psi}\bra {\psi^\perp}$,
as:
\begin{align}
&\rho_1(t) =e^{-\Ga {\cal D}_L t} \rho_1(0) = \ket{\psi} \bra {\psi} + O(e^{-\Ga t}).
\end{align}
Note that the a similar solution can be written  also for $\Ga \gg ||H_s||$, i.e.  when the coherent evolution can be considered as a small perturbation.
From the above it follows that the targeting of the pure states at the edges
\begin{align*}
&\rho_{L,R} =\ket{\psi_{L,R}} \bra {\psi_{L,R}}\\
&\bra {\psi_{\gamma}}  =\left(\cos \frac{\th_\gamma}{2}  ,\quad e^{-i \vfi_\gamma} \, \sin \frac{\th_\gamma}{2} \right), \;\; \ga=L, R ,
\end{align*}
can be achieved by a dissipator ${\cal D}$ in (\ref{eq:LME}) with Lindblad operators
\begin{align}
&L_1 =\ket{\psi_{L}} \bra {\psi_{L}^\perp} \otimes I^{\otimes_{N+1}} \label{eq:L1}, \\
&L_2 = I^{\otimes_{N+1}} \otimes \ket{\psi_{R}} \bra {\psi_{R}^\perp}\label{eq:L2},
\end{align}
acting at sites $0$, $N+1$, respectively. It can be proved that the steady state, $\rho_{NESS}(\Ga)=\lim_{t\rightarrow \infty} \rho(\Ga,t)$, of (\ref{eq:LME})
is unique for any given choice of the targeted polarizations and in the Zeno limit we have
 \begin{align}
&\lim_{\Ga\rightarrow \infty}  \rho_{NESS}(\Ga) =  \rho_{NESS}^{Zeno} = \rho_L \otimes R \otimes \rho_R.
\label{eq:Zeno}
\end{align}

We also recall (see \cite{2018ZenoDynamics} for details) that at time scale $t=O(1)$ and large  $\Ga$, the density matrix $\rho(t)$  acquires the approximate form (\ref{eq:Zeno}) with  $R\equiv R(t \Gamma)=R(\tau)$, evolving  in time according to
\begin{align}
&\frac{\partial R(\tau)} {\partial \tau} = - i [h_D,R] +\frac{1}{ \Ga} {\cal D}_{eff}[R] \label{eq:LMEeff}
\end{align}
where $\cal{D}_{eff}$ is an effective dissipator and $h_D = {h_D}^\dagger$ is the so-called dissipation-projected Hamiltonian \cite{2014ZanardiVenuti} given in Eq.(1) of the main text.
From Eqs. (\ref{eq:Zeno}), (\ref{eq:LMEeff}) it follows that $[h_D,R]=0$ and $R$ is given by
\begin{align}
&R(\tau) = \sum_\al P_{\al}(\tau) \ket{\al} \bra{\al}+ O(\Ga^{-1}),
\end{align}
with $\ket{\al}$ denoting $h_D$ eigenvectors. Moreover, the coefficients $P_{\al}(\tau)$
evolve adiabatically under the  perturbative effective dissipator ${\cal D}_{eff}$ in (\ref{eq:LMEeff}). The adiabatic evolution is given by the  Markov process discussed in sec.~\ref{S-III}.
Remarkably, it was shown in \cite{2020Dirichlet-PRE} that the Zeno NESS (\ref{eq:Zeno}) is robust against possible asymmetries of the dissipation between left and right boundaries, i.e. the
asymmetric rescaling of the Lindblad operators $L_1 \rightarrow \delta_1 L_1, L_2 \rightarrow \delta_2  L_2 $ in (\ref{eq:LME}), with any finite nonzero
$\delta_1,\delta_2$ values, yields the same limiting result (\ref{eq:Zeno}).

\section{Symmetry properties of the Lindblad equation}
\label{S-Ib}
In the following we investigate symmetry properties of the NESS of the  Lindblad equation (\ref{eq:LME}) with  targeted polarizations lying in $XY$-plane, $\th_L=\th_R=\pi/2$, $\vfi_L=0$, and arbitrary $\vfi_R$. For this we introduce the operator  $U_\vfi$ performing a simultaneous rotation of all spins around the $z$ axis and the parity operator $\mathcal{P}$, respectively defined as
\begin{align*}
U_\vfi = \prod_{k=0}^{N+1} e^{i \frac{\vfi}{2} \si_k^z}, \quad\quad \mathcal{P}:= S_n \rightarrow S_{N-n+1}, \quad\ n=0,1,...N+1.
\end{align*}
Notice that the bulk Hamiltonian $H_s$ commute with both operators $U_\vfi$ and $\mathcal{P}$ and hence also with their product $U_\vfi \mathcal{P} \equiv \mathcal{U}_\vfi$, i.e.  $[H, \mathcal{U}_\vfi]=0$. On the other hand the Lindblad operators $L_1,L_2$ from (\ref{eq:L1}), (\ref{eq:L2}), under the action of $\mathcal{U}_{\vfi_R}$ transform as:
\begin{align*}
&\mathcal{U}_{\vfi_R} \  L_2  \ \mathcal{U}_{\vfi_R}^\dagger =  L_1,\\
& \mathcal{U}_{\vfi_R} \ L_1 \ {\mathcal{U}_{\vfi_R}}^\dagger =
e^{i \vfi_R \si_{N+1}^z} L_2 e^{-i \vfi_R \si_{N+1}^z}= \left. L_2 \right|_{\vfi_R \rightarrow -\vfi_R},
\end{align*}

i.e. one gets the same physical setup but with the opposite boundary gradient at the right boundary. Indeed, it is straightforward to verify that LME (\ref{eq:LME}) under the $\mathcal{U}$ transformation is mapped onto the same LME but with opposite boundary gradient, $\vfi_R \rightarrow -\vfi_R$, with  the corresponding $\rho$ matrix  transformed as
\begin{align}
&\rho (-\vfi_R) = \mathcal{U}_{\vfi_R} \  \rho(\vfi_R) \ \mathcal{U}^\dagger_{\vfi_R}.
\label{eq:NESStransform}
\end{align}
The uniqueness of the NESS for any choice of the boundary parameters, implies that the above transformation is a map between NESS of  opposite boundary gradients. On the other hand, taking into account that $M_{+} + M_{-}=N+1$ we have,  from the $\vfi_R$ angles parametrized by the integer $M_{+}$ in Eq.(\ref{eq:Split}),  that:
\begin{align}
&\vfi_R(M_{-}) \equiv (N+1 -2 M_{-})\ga = (N+1- 2 (N+1-M_{+}))\ga= -(N+1 -2 M_{+})\ga = -\vfi_R(M_{+}),
\end{align}
thus the transformation (\ref{eq:NESStransform}) allows to get the $M_{-}-$NESS from the $M_{+}$ one as:
\begin{align}
&\rho_{NESS} (\vfi_R(M_{-})) = \mathcal{U}_{\vfi_R(M_{+})}  \  \rho_{NESS}(\vfi_R(M_{+})) \  \mathcal{U}^\dagger_{\vfi_R(M_{+})}.
\label{eq:NESStransform1}
\end{align}

 Notably, the steady state current under the transformation $\vfi_R \rightarrow -\vfi_R$  changes its sign: $\langle j^z (M_{-})\rangle = -
\langle j^z  (M_{+}) \rangle$, in agreement with what observed in Fig.~2 and Fig.~3 of the main text.

\section{Invariant subspace $W_M$ of $h_D$ at phantom Bethe manifolds}
\label{S-II}

To investigate the invariant phantom Bethe manifolds it is convenient to introduce the following   parametrization for an arbitrary pure qubit state
\begin{align}
&\ket{f,F}=\frac{1}{\sqrt{1+e^{2f}}}\binom {1}{ e^{f +i F}}, \label{DefQubit}\\
&\ket{F}=\frac{1}{\sqrt{2}} \binom {1}{ e^{i F}},
\end{align}
where  $f,F$ are real numbers. Notice that the state $\ket{f,F}$ describes a spin $1/2$ pointing in the direction
$( \langle \si^x \rangle,  \langle \si^y \rangle,  \langle \si^z \rangle)= (\sin \th \cos F,\sin \th \sin F, \cos \th) $ with $\tan( \th/2)= e ^{f}$, thus $-\infty<f<\infty $, parametrizes the polar angle  while $0\leq F<2 \pi$ parametrizes the azimuthal angle (or phase factor) of the usual polar coordinate system. The state $\ket{F}$ instead,  describes a fully polarized spin $1/2$ lying in $XY-$plane. Here for simplicity we consider the condition (\ref{eq:Split}) at $\th=\pi/2$ (dissipation baths polarizations in the XY plane) to  prove the following Theorem.

\textbf{Theorem} The set of states
\begin{align}
&\kket{n_1,n_2, \ldots n_M}  =  \bigotimes_{k=1}^{n_1-1} \ket{ k \vfi}  \bigotimes_{k=n_1}^{n_2-1} \ket{(k-2) \vfi}
\ldots  \bigotimes_{k=n_M}^{N} \ket{ (k-2M) \vfi},
\label{eq:Gprimebasis}\\
 &1\leq n_1 <n_2 < \ldots < n_M \leq N+1,
\end{align}
form an invariant subspace $W_M^{+}$ of $h_D$ (\ref{eq:hD}) satisfying Eq. (\ref{eq:Split}) with fixed $M$ and $\vfi_L=0$.
\bigskip

\textit{Remark.--} In (\ref{eq:Gprimebasis}) $n_1=1$ and $n_M=N+1$ denote virtual kinks at outer links of the chain $(0,1)$ and $(N,N+1)$: $n_1=1$ and
$n_1>1$ correspond to  first qubit in (\ref{eq:Gprimebasis}) being $\ket{-\vfi}$ and  $\ket{\vfi}$ respectively.
Likewise, $n_M=N+1$ and
$n_M<N+1$ correspond to  the last qubit of the form $\ket{(N-2M+2)\vfi}$ and
$\ket{(N-2M)\vfi}$ respectively.
Introducing the notation
\begin{align}
&  \ket{F_1,F_2, \ldots F_m}=2^{-m/2} \bigotimes_{k=1}^{m} \binom {1}{ e^{i F_k}}, \label{Frepresentation}
\\
&F^\perp= F+ \pi; \quad \ket{F^\perp} = \si^z \ket{F}; \quad \braket{F^\perp }{F}=0,
\end{align}
the operator $\tilde{h}_D  = h_D - (N-1)\De I$  can be written as sum of local terms
\begin{align}
&\tih_D=h_{12}+ h_{23} + \ldots + h_{N-1,N} + h_l \otimes I^{\otimes_{N-1}} + I^{\otimes_{N-1}}\otimes h_r,
\end{align}
where $h=\si^x \otimes \si^x + \si^y \otimes \si^y + \De (\si^z \otimes \si^z - I \otimes I)$
is the local energy-density of the $XXZ$ spin chain.

A key point in the proof  is the  divergence condition:
\begin{align}
&h  \ket{F,F+\ga}= -\ka \ket{F^\perp,F+\ga} + \ka  \ket{F,(F+\ga)^\perp},
\label{div}\\
&\ka =  i \sin \ga, \nonumber\\
&\De = \cos \ga.
\end{align}
Using  (\ref{div}) one obtains the following  useful relations:
\begin{align}
&(h_{12}+h_{23})  \ket{F,F+\ga,F} = -\ka  \ket{F^\perp,F+\ga,F} - \ka  \ket{F,F+\ga,F^\perp} +a_{0}  \ket{F,F+\ga,F} + a_{+}  \ket{F,F-\ga,F}, \label{X123}\\
&(h_{12}+h_{23})  \ket{F,F-\ga,F} = \ka  \ket{F^\perp,F-\ga,F} + \ka  \ket{F,F-\ga,F^\perp} + a_{0}  \ket{F,F-\ga,F} + a_{-}  \ket{F,F+\ga,F}, \label{Y123}\\
& a_0 = -2 \De, \quad a_{\pm} = 2 e^{\pm i \ga}.
\end{align}
Due to  (\ref{div}),(\ref{X123}), (\ref{Y123}), any factorized state of the form:
\begin{align}
& \Psi = \otimes_{k=1}^N  \ket{0,F_k}, \quad \text{with} \quad F_{k+1}-F_{k} = \pm \ga,
\label{Psi}
\end{align}
under the action of the operator $\sum_{n=1}^{N-1} h_{n,n+1}$ will transform into a sum of terms $\Psi_\al$ of the same  type of (\ref{Psi}), i.e.
$\Psi_\al=\otimes_{k=1}^N  \ket{F_{k,\al}}, \text{with }  F_{k+1,\al}-F_{k,\al} = \pm \ga,$
plus two extra terms :
\begin{align*}
&\sum_{n=1}^{N-1} h_{n,n+1} \Psi = \sum_{\al}  C_\al \ket{F_1,F_{2,\al}F_{3,\al} \ldots  F_{N-1,\al},F_N} + A_1 \ket {F_1^\perp, F_2, \ldots  ,F_N} + A_N \ket {F_1, F_2, \ldots  F_{N-1},F_N^\perp},  \\
&F_{2\al}- F_1 = \pm \ga, \quad F_{(k+1),\al} - F_{k,\al} = \pm \ga; \quad F_N -  F_{N-1,\al} = \pm \ga,
\end{align*}
with $A_1=  \ka \  sign((F_1-F_2)/\ga)$ and $A_N=  \ka  \ sign((F_N-F_{N-1})/\ga)$.
All basis vector of $W_M$ are of the form (\ref{Frepresentation})
with $F_1= \pm \ga$ , $F_N = F_c(M) \pm \ga$.
By acting with $\tilde{h}_D$ on the basis elements $\kket{n_1 n_2 \ldots n_M}$,
we generate other terms of the same basis $\kket{m_1 m_2 \ldots m_M}\equiv \ket{F_1, \ldots F_N}$
and extra terms with "impurities" at site $1$ and site $N$:

\begin{align*}
&\tih_D  \kket{n_1 n_2 \ldots n_M} = \ldots + A_1 \ket {F_1^\perp, F_2, \ldots  }  +  h_l \otimes I^{\otimes_{N-1}} \ket {F_1^\perp, F_2, \ldots  } + \\
&+ A_N \ket {\ldots F_{N-1},F_N^\perp} +  \left(I^{\otimes_{N-1}}\otimes h_r \right)  \ket {\ldots ,F_{N-1},F_N^\perp}.
\end{align*}
Since $A_1= sign((F_1 - F_2)/\ga) \ka$ and $F_1 = \pm \ga$, we have that the following four
$A_1 \ket{F_1^\perp,F_2\ldots}$  terms may arise:
\begin{align}
& A_1 \ket {F_1^\perp, F_2, \ldots  }= \ka \ket {-\ga^\perp,-2\ga,\ldots}, \quad - \ka \ket {\ga^\perp,2\ga, \ldots } \label{NonTrivial}, \\
& A_1 \ket {F_1^\perp, F_2, \ldots  }= \ka \ket {\ga^\perp,0,\ldots}, \quad - \ka \ket {-\ga^\perp,0, \ldots}. \label{Trivial}
\end{align}
Terms of type (\ref{NonTrivial}) cancel with respective terms from $h_l \otimes I^{\otimes_{N-1}}  \ket {F_1,F_2 \ldots}$, using the relations
\begin{align}
&h_l \ket{\pm \ga} =\pm \ka  \ket{\pm \ga^\perp} + \De   \ket{\pm \ga},
\label{hl1}
\end{align}
and similarly, one get  cancelation of the terms of type (\ref{Trivial}) by means of the relations
\begin{align}
&h_l \ket{\pm \ga} =\mp \ka  \ket{\pm \ga^\perp} - \De   \ket{\pm \ga} + 2 e^{\pm i \ga}   \ket{\mp \ga}.
\label{hl2}
\end{align}
Proceedings in the same manner for the arising four terms of type $A_N \ket {\ldots F_{N-1},F_N^\perp}$:
\begin{align}
& A_N \ket {\ldots F_{N-1},F_N^\perp}  =  -\ka  \ket {\ldots, \Phi_c +2 \ga,  (\Phi_c + \ga)^\perp}, \quad   \ka  \ket {\ldots, \Phi_c -2 \ga,  (\Phi_c - \ga)^\perp},   \label{NTr}\\
& A_N \ket {\ldots F_{N-1},F_N^\perp}  =  \ka  \ket {\ldots, \Phi_c,  (\Phi_c + \ga)^\perp}, \quad   -\ka  \ket {\ldots, \Phi_c,  (\Phi_c - \ga)^\perp},   \label{Tr}
\end{align}
with
\begin{align}
&\Phi_c(M) =(N+1-2M)\ga, \label{eq:Phi(M)}
\end{align}
one finds that terms (\ref{NTr}),(\ref{Tr}) cancel with respective terms $I^{\otimes_{N-1}} \otimes h_r \ket{\ldots  F_{N-1},F_N}$ using relations
\begin{align}
&h_r \ket{\Phi_c \pm \ga} =\pm \ka  \ket{(\Phi_c \pm\ga)^\perp} + \De   \ket{\Phi_c \pm \ga}
\label{hr1}, \\
&h_r \ket{\Phi_c \pm \ga} =\mp \ka  \ket{(\Phi_c \pm\ga)^\perp} - \De   \ket{\Phi_c \pm \ga}  + 2 e^{\pm i \ga}   \ket{\Phi_c \mp \ga},
\label{hr2}
\end{align}
respectively (note that relations  (\ref{hr1}), (\ref{hr2}) differ from  (\ref{hl1}), (\ref{hl2}) simply by the increase of the angle by $\Phi_c(M)$). From this we conclude  that the action of $h_D$ on any basis vector of $W_M$ produces states that still belong to $W_M$, thus proving the Theorem (QED). We close this section with the following remarks.

\textbf{Remark 1.--}
It can be shown that the repeated action of $h_D$ on any single basis vector generates the whole subspace $W_M$. Moreover, it follows that all eigenvectors $\ket{\al_+} \in W_M$ of $h_D$ have nonzero overlaps with all vectors $\kket{n_1 \ldots n_M}$ of $W_M$ basis. The total number of states in $W_M$ in (\ref{eq:Gprimebasis}) can be calculated combinatorially as:
\begin{align*}
& \dim \ W(M) = \binom{N+1}{M} = \binom{N}{M} +  \binom{N}{M-1}.
\end{align*}

\textbf{Remark 2.--}
The subspace $W_M$ is a part of a larger invariant subspace $G_M^{+}$ described in
\cite{PhantomLong}. $G_M^{+}$ contains all vectors of $W_M$ plus additional linearly independent
vectors, all with qualitatively the same chirality features. The total number of vectors in $G_M^{+}$
is
\begin{align*}
& \dim \ G_M^{+} =  \binom{N}{M} +  \binom{N}{M-1} +   \binom{N}{M-2}+ \ldots +   \binom{N}{0},
\end{align*}
see \cite{PhantomLong}.
The appearance of smaller invariant subspace $W(M)$ within $G_M^{+}$ is a consequence of a special
form of the boundary fields in the dissipation-projected Hamiltonian (\ref{eq:hD}).

\section{Properties of Markov process (\ref{eq:MP}) at points $\Phi_c(M)$}
\label{S-III}
The effective Markov evolution for the probabilities $P_\al$ of the occupation of state $\al$ at time $\tau$:
\begin{align}
&\frac{\partial P_\al(\tau)} {\partial \tau} = \sum_{\be \neq \al} w_{\be \al} P_\be(\tau) - \sum_{\be \neq \al} w_{\al \be} P_\al(\tau)
\label{eq:MPapp}
\end{align}
is related to a reduced density matrix dynamics $R(\tau)$, near the quantum Zeno limit, given by
\begin{align*}
& R(\tau) \approx \sum_{\al=1}^{dim {\cal H}} P_\al(\tau) \ket{\al} \bra{\al},
\end{align*}
where $dim {\cal H}=2^N$ is the dimension of the Hilbert space, and  $\ket{\al}$ are eigenvectors of $h_D$.
The rates $w_{\al \be}\equiv w_{\al \rightarrow \be}$ are given by
\cite{2018ZenoDynamics}:
\begin{align}
&w_{\al \be} = |\bra{\be} g_L \ket {\al} |^2 +  |\bra{\be} g_R \ket {\al} |^2,
\label{eq:MPratesApp}
\end{align}
where $g_L,g_R$ are some locally acting operators\cite{2018ZenoDynamics}.
On phantom Bethe manifolds parametrized by the integer number $M$,
the existence of invariant subspaces of $h_D$:  $G_M^{+},G_M^{-}$ and $W_M \in G_M^{+}$ allows to split the $h_D$ eigenvectors $\ket{\al}$ into three blocks, as

\begin{align}
& R(t) \approx \sum_{\al_{+}^{(1)}=1}^{dim (W_M)} P(\al_+^{(1)},t) \ket{\al_+^{(1)}} \bra{\al_+^{(1)}} +\nonumber \\
&+\sum_{\al_{+}^{(2)}=1}^{dim (G_M^+)-dim (W_M)} P(\al_+^{(2)},t) \ket{\al_+^{(2)}} \bra{\al_+^{(2)}} +  \label{R(t)}\\
&+ \sum_{\al_{-}=1}^{dim(G_M^{-})} P(\al_-,t) \ket{\al_-} \bra{\al_-}, \nonumber
\end{align}
corresponding to  respective $h_D$ eigenvalues $\{ \la_{\al_{+}^{(1)}}\}, \ \{ \la_{\al_{+}^{(2)}}\}$ and $\{ \la_{\al_{-}}\}$.
 In the Eq (\ref{R(t)}), the first two block contributions share the same chirality (indicated with $"+"$ subscript).
It was proved in \cite{PhantomLong} that states in  $G_M^{+},G_M^{-}$ are mutually orthogonal.
If the sets $\{ \la_{\al_{+}^{(1)}}\}$, $ \{ \la_{\al_{+}^{(2)}}\}$ have no intersection,
\begin{align}
& \{ \la_{\al_{+}^{(1)}} \} \cap \{ \la_{\al_{+}^{(2)}} \} =  \varnothing, \label{NoIntersec}
\end{align}
then also
states $\ket{\al_{+}^{(1)}}$ , $\ket{\al_{+}^{(2)}}$ are also orthogonal.

We will be interested in cumulative probability current towards set of states in $W_M$
from other blocks, given by
\begin{align}
& J_{\al_{-},\al_{+}^{(2)} \rightarrow \al_{+}^{(1)}} (t) = \sum_{\al_{+}^{(1)}: \ \ket{\al_{+}^{(1)}} \in W_M}  \quad \sum_{\be: \ \ket{\be} \notin W_M} \left(
P(\be,t) w_{\be \al_{+}^{(1)}} - P(\al_{+}^{(1)},t) w_{\al_{+}^{(1)} \be}.
\right)
\label{Jprob}
\end{align}
First, notice that all  basis vectors $\kket{n_1,n_2, \ldots n_M}$ of $W_M$
are eigenvectors of operators $g_L, g_R$ in (\ref{eq:MPratesApp}):
\begin{align*}
&g_L \kket{n_1,n_2, \ldots n_M}  = (2 \de_{n_1,1}-1) i \sin \vfi  \kket{n_1,n_2, \ldots n_M}, \\
&g_R \kket{n_1,n_2, \ldots n_M}  = (1-2 \de_{n_M,N+1}) i \sin \vfi  \kket{n_1,n_2, \ldots n_M},
\end{align*}
entailing all
\begin{align}
& w_{\al_{+}^{(1)} \be} = 0; \quad \ket{\al_{+}^{(1)}} \in  W_M, \quad \ket{\be} \notin  W_M, \label{wout}
\end{align}
provided (\ref{NoIntersec}).
On the other hand, at least some of the $w_{\be \al_{+}^{(1)}} $ rates in  (\ref{Jprob}) are strictly positive,
since  $W_M$ basis states $\bbra{n_1 n_2 \ldots n_M}$ are not left eigenvectors
of $g_L,g_R$.
Consequently,
\begin{align}
& w_{\be \al_{+}^{(1)}} > 0;   \quad  \ket{\al_{+}^{(1)}} \in  W_M, \quad \ket{\be} \notin  W_M, \mbox {for some } \be. \label{win}
\end{align}
Eqs(\ref{wout}),  (\ref{win}) show that one has strictly positive probability current (\ref{Jprob})
at all times, leading to 
gradual increase of cumulative population of
 states with the same chirality
$\sum P( \al_{+}^{(1)},t)$
 with time, and depopulation of all other states. In the long time limit, only contributions from the invariant subspace $W_M$ in (\ref{R(t)}) remain,
leading to
\begin{align*}
& R(t \rightarrow \infty) = \sum_{\al_{+}^{(1)}=1}^{dim (W_M)} P(\al_{+}^{(1)}) \ket{\al_{+}^{(1)}} \bra{\al_{+}^{(1)}} + O(\Ga^{-1}).
\end{align*}

\end{document}